\documentclass[prl,twocolumn,showpacs]{revtex4}
\usepackage{epsfig}
\usepackage{amsbsy}
\usepackage{amsmath}
\usepackage{latexsym}

\begin{document}

\title{Bipartite entanglement and localization of one-particle states}
\author{Haibin Li$^{(1,2)}$, Xiaoguang Wang$^{(1,3)}$, and Bambi Hu$^{(1,4)}$}
\affiliation{1. Department of Physics and Center for Nonlinear Studies, Hong Kong Baptist University, Hong Kong, China.}
\affiliation{2. Zhejiang Institute of Modern Physics, Zhejiang University, Hangzhou 310027, China.}
\affiliation{3. Department of Physics and Australian Centre of Excellence for Quantum
Computer Technology, \\
Macquarie University, Sydney, New South Wales 2109, Australia.}
\affiliation{4. Department of Physics, University of Houston, Houston, Texas 77204-5005, USA.}

\date{\today}
\begin{abstract}
We study bipartite entanglement in a general one-particle state, and find that 
the linear entropy, quantifying the bipartite entanglement, is directly connected to the 
paricitpation ratio, charaterizing the state localization. 
The more extended the state is, the more entangled the state.
We apply the general formalism 
to investigate ground-state and dynamical 
properties of entanglement in the one-dimensional 
Harper model.
\end{abstract}
\maketitle

Entanglement is a unique phenomenon of quantum systems that does not exist
classically. It has attracted more interest due to its potential applications 
in quantum communication and information processing~\cite{Nie00} such as quantum
teleportation~\cite{Ben1}, superdense coding~\cite{Ben2}, quantum key
distribution~\cite{Ekert}, and telecoloning~\cite{Mur}.
On the other hand, entanglement has been proved to be playing an important role
in condensed matter physics. There are many studies on entanglement 
in the 
Heisenberg spin models~\cite{Heisen1,Heisen2,Wang1}, Ising models in a transverse magnetic
field~\cite{Connor,Gun}, and related itinerant fermionic systems~\cite{Fermion}.
In the context of quantum phase transition, entanglment is also an indicator of 
the transition which can not be captured by common statistical physics~\cite{Osb, Ost}.

Recently, {\it pairwise entanglement} sharing in one-particle states was studied~\cite{Lak03}
using the concurrence~\cite{Conc} in the Harper model~\cite{Harper}. Here, we study
another type of entanglement of one-particle states, the {\it bipartite entanglement}, which refers
to entanglement between two subsystems when a whole system is divided into two parts. 
We reveal that the average bipartite entanglement directly connects to state localization.

The one-particle states permeate many physics systems. For examples, 
for one electron moving on a substrate potential, the eigenfunctins are one-particle states. 
In quantum spin chain models with only one
spin up (down) and all other spins down (up), the eigenfuntions of the model are
one-magnon states.

We consider a system containing $N$ two-level systems (qubits) with $|0\rangle$ being the excited state
and $|1\rangle$ the ground state. A general one-particle state is then written as
\begin{align}
|\Psi\rangle=&\psi_{1}|1000\ldots 0\rangle +\psi _{2}|0100\ldots 0\rangle \nonumber\\
&+\ldots +\psi_{N}|0000\ldots 1\rangle \label{general}
\end{align}
Here, $\{|\psi _{n}|^2\}$ is a probability distribution, satisfying the 
normalization condition 
$
\sum_{n=1}^N|\psi_{n}|^2=1. 
$
When $|\psi_n|=1/\sqrt{N}$, state $|\Psi\rangle$ reduces to the W state~\cite{W,Wang1},
one representative state in quantum information theory.

We now consider bipartite entanglement between a block of $L$ qubits and the rest $N-L$ qubits.
Bipartite entanglement of a pure state can be measured 
by the linear entropy of reduced density 
matrices~\cite{Ben3}.
\begin{equation}
E(|\Psi\rangle)=1-\text{Tr}(\rho_{i}),\quad  i\in\{1,2\}
\label{linear}
\end{equation}%
where $\rho_{i}$ is the reduced density matrix for subsystem $i$.

To calcualte bipartite entanglement, we first consider a simple situation, namely, 
the entanglement between the first qubit and the rest $N-1$ qubits.
The one-particle state can be written in the following form:
\begin{equation}
|\Psi \rangle=\psi _{1}|1\rangle \otimes |\alpha\rangle+
\sqrt{\sum_{n=2}^N|\psi_{n}|^2}|0\rangle \otimes |\beta\rangle 
\end{equation}
where 
\begin{align}
|\alpha\rangle =&|00\ldots 0\rangle, \nonumber\\
|\beta \rangle =&\frac{1}{\sqrt{\sum_{n=2}^N|\psi_{n}|^2}}
(\psi_{2}|100\ldots 0\rangle +\psi _{3}|010\ldots 0\rangle\nonumber\\
&+\ldots +\psi_{N}|00\ldots1\rangle) 
\end{align}
are two orthonormal states of $N-1$ qubits.
Then the reduced density matrix for qubit 1 is easily found to be
\begin{equation}
\rho_1=\left( 
\begin{array}{cc}
1-|\psi_{1}|^{2} & 0 \\ 
0 & |\psi_{1}|^{2}%
\end{array}%
\right) \label{reduce}
\end{equation}
in the basis $\{|0\rangle,|1\rangle\}$.
Therefore, from Eqs.~(\ref{linear}) and (\ref{reduce}), 
the linear entropy of qubit 1 is obtained as 
$
E_{1,N-1}^{(1)}=2(|\psi_{1}|^{2}-|\psi_{1}|^{4}),
$
where the superscript denotes the first qubit. In the same way, 
we may find the linear entropy for the $n$-th qubit as
\begin{equation}
E_{1,N-1}^{(n)}=2(|\psi _{n}|^{2}-|\psi _{n}|^{4}),
\end{equation}
quantifying the degree of bipartite entanglement 
between $n$-th qubit and the rest.

Making an average of entanglement over all qubits, we obtain
\begin{align}
E_{1,N-1}=\langle E_{1,N-1}^{(n)}\rangle =&\frac{2}{N}\sum_{n=1}^{N}(|\psi_{n}|^{2}-
|\psi_{n}|^{4}) \nonumber\\
=&\frac{2}{N}\left(1-\sum_{n=1}^{N}|\psi_{n}|^{4}\right)=\frac{2}{N}E_{s},
\end{align}
where $E_{s}=(1-\sum_{n=1}^N|\psi_{n}|^{4})$ 
is the quantum state linear entropy~\cite{Georgeot00}  for state $|\Psi\rangle$. 

We now consider more general situations, namely, the bipartite entanglement
between a block of $L$ qubits and the rest of the system.
We pick up $L$ qubits $\{1^{{\prime}},2^{{\prime}},\ldots,L^{{\prime }}\}$ from $N$ qubits,
and there are totally $C_N^L$ options. 
In the same way as above, we can calculate the linear entropy of $L$ qubits as
\begin{equation}
E_{L,N-L}^{\{1^\prime,\ldots,L^\prime\}}=
2(|\psi_{1^{{\prime }}}|^{2}+\ldots+|\psi _{L^{^{\prime }}}|^{2})
-2(|\psi_{1^{{\prime }}}|^{2}+\ldots+|\psi _{L^{^{\prime }}}|^{2})^2.
\end{equation}

Now, we make an average of the linear entropy $E_{L,N-L}^{\{1^\prime,\ldots,L^\prime\}}$ over
the $C_N^L$ options. Formally, the average entropy is given by
\begin{align}
E_{L,N-L}=&\langle 
E_{L,N-L}^{\{1^\prime,\ldots,L^\prime\}}\rangle=\Big[\sum_{\{1^\prime,\ldots,L^\prime\}}
\frac{2}{C_N^L}(|\psi_{1^{{\prime }}}|^{2}+\ldots+|\psi _{L^{^{\prime }}}|^{2})\nonumber\\
-&\frac{2}{C_N^L}(|\psi _{1^{{\prime }}}|^{2}+\ldots+|\psi _{L^{^{\prime 
}}}|^{2})^2\Big].\label{main}
\end{align}
The first term in the summation of the above equation can be easily evaluated as
\begin{equation}
\frac{2}{C_N^L}\sum_{\{1^\prime,\ldots,L^\prime\}}(|\psi _{1^{{\prime }}}|^{2}+|\psi _{2^{{\prime
}}}|^{2}+\ldots+|\psi_{L^{{\prime }}}|^{2})=\frac{2L}{N}. \label{main1}
\end{equation}

Next, we evaluate the second term of the summation in Eq.~(\ref{main}). The following relation 
\begin{equation}
\sum_{n>m}2|\psi_n|^2|\psi_m|^2=1-\sum_{n=1}^N |\psi_n|^4.
\end{equation}
is useful, resulting from the normalization condition of state $|\Psi\rangle$. 
Note that the summation on the left hand of the above 
equation contains $C_N^2$ terms.
By applying the above relation, the second term of the summation in Eq.~(\ref{main}) is obtained as
\begin{equation}
-\frac{2L}{N}\sum_{n=1}^N|\psi_n|^4-\frac{2 C_L^2}{C_N^2}\left(1-\sum_{n=1}^N|\psi_n|^4\right). 
\label{main2}
\end{equation}
Substituting Eqs.~(\ref{main1}) and (\ref{main2}) to Eq.~(\ref{main}) leads to
\begin{equation}
E_{L,N-L}=2\left(\frac{L}{N}-\frac{C_L^2}{C_N^2}\right)E_s=\frac{2L(N-L)}{N(N-1)}E_s. 
\label{mainresult1}
\end{equation}
As we expected, the average linear entropy is invariant under the transformation
$L\rightarrow N-L$. 

The above result (\ref{mainresult1}) shows that the average bipartite entanglement between a block of 
$L$ qubits
and the rest $N-L$ qubits is proportional to the quantum state linear entropy $E_s$.
For the W state, the quantum state linear entropy $E_s=1-1/N$, and thus
the average linear entropy becomes
\begin{equation}
E_{L,N-L}=2L(N-L)/N^2.
\end{equation}
If $N$ is even being fixed and $L=N/2$, 
the linear entropy $E_{L,N-L}=1/2$ becomes maximal, 
equal to the amount of bipartite entanglement of a Bell state. 

There exists a close relation between the average {\it pairwise entanglement} and state localization 
for one-particle states as discussed in
Ref.~\cite{Lak03}. We now study relations between {\it bipartite entanglement} and state localization. 
The degree of localization can be studied
by a simple quantity, the participation ratio defined by
\begin{equation}
p=\frac{1}{N\sum_{n=1}^N|\psi_n|^4}.\label{ppp}
\end{equation}
Then, from the above equation, the quantum state linear entropy can be written as
\begin{equation}
E_s=1-\frac{1}{Np}.\label{Esp}
\end{equation}
Subsitituting Eq.~(\ref{Esp}) to Eq.~(\ref{mainresult1}) leads to
\begin{equation}
E_{L,N-L}=\frac{2L(N-L)}{N(N-1)}\left(1-\frac{1}{Np}\right), \label{mainresult2}
\end{equation}
which builds a direct connection between bipartite entanglement and state localization.
It is evident that the larger $p$ is, the larger the linear entropy. In other words, the more extended
the one-particle state is, the more entangled the state.

In a recent relevant work~\cite{BambiLiWang1}, we find that there is a relation between the average square of  concurrence and the
participation ratio given by
\begin{equation}
\langle C^2\rangle=\frac{4}{N(N-1)}\left(1-\frac{1}{Np}\right). \label{result}
\end{equation}
Then, from Eqs.~(\ref{mainresult2}) and (\ref{result}), we obtain a connection between bipartite entanglement and pairwise entanglement
given by
\begin{equation}
E_{L,N-L}=\frac{L(N-L)}{2}\langle C^2\rangle.
\end{equation}
Therefore, for one-particle state, the average linear entropy, the average square of concurence, 
and the participation ratio are interrelated together by simple relations. Next, we consider an example
of one-particle states, and study quantum entanglement and its relations to 
state localization induced by on-site potentials 
in the quasiperiodic one-dimensional Harper model~\cite{Harper}.

The Hamiltonian describing electrons hopping in a one-dimentional lattice can be written 
as~\cite{Harper} 
\begin{equation}
H=\sum_{n=1}^N\frac{1}{2}(c_{n}^{\dagger}c_{n+1}+c_{n+1}^\dagger c_n)+V_nc_{n}^{\dagger}c_{n},
\end{equation}%
where $c_{n}^{\dagger}$ and $c_{n}$ are the creation and annihilation operators, 
respectively, and $V_n$ is the on-site potential. This Hamiltonian describes 
electrons moving on a substrate potential. 
The different forms of on-site potential $V_n$ lead to different 
behaviors of electrons. 

For the Harper model, the on-site potential is given by
\begin{equation}
V_n=\lambda \cos (2\pi n\sigma )
\end{equation}
where $\sigma $ determines the period of potential and $\lambda$ is the
amplitude of the potential. 
It is well known that the dynamics of the Harper model is
characterized by parameter $\lambda$~\cite{Gei}. If $\lambda <1,$ the
electron is in a quasiballistic extended state, but in a localized
state when $\lambda >1.$ The critical Harper model corresponds to $\lambda =1$,
where the spectrum is a Cantor set.

We consider one-electron states in the Harper model.
For studying entanglement in this fermionic system, we adopt
the approach given in Ref.~\cite{Fermion}, namely, we first mapping
electronic states to qubit states and study entanglement of fermions by
calculating entanglement of qubits. After the mapping, the one-particle wavefunction 
of Harper model can be formally written in the form of a general state given by Eq.~(\ref{general}). 
Then, we may apply the general result (\ref{mainresult1}) to calculate bipartite
entanglement.

\begin{figure}
\includegraphics[width=0.40\textwidth]{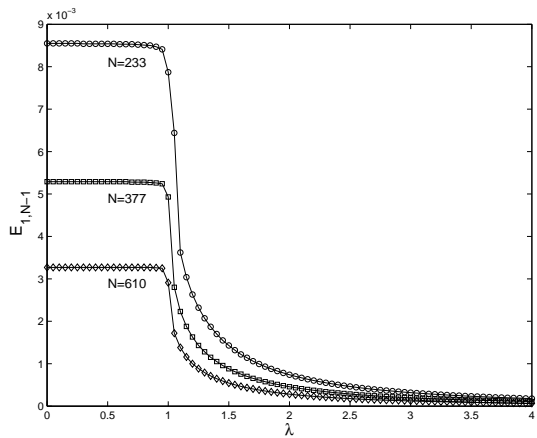}
\caption{The average linear entropy as a function of $\lambda$ for differnet 
lengths of the chain. The parameter $\sigma$ is $F(n-1)/F(n)$ with $F(n)=N$, where
$\{F(n)\}$ is the Fibonacci sequence.}
\end{figure}

For convenience, we examine the average bipartite entanglment 
between one local fermionic mode (LFM) and the rest for the ground state of the system. 
The LFMs refer to sites which can be either empty
or occupied by an electron~\cite{Bravyi}. 
Figure 1 gives results of the linear entropy for different lengths of system. 
We notice that when the amplitude of the on-site potential increases but does not 
reach the value $\lambda=1$, the average entanglement $E_{1,N-1}$ 
is almost keeping a constant value, varying a very little bit. 
The entanglement is not destroyed  by the external potential in this
region. However, when $\lambda =1$, $E_{1,N-1}$ has a sudden jump down 
to a value close to zero.
For larger $\lambda$, the ground state is almost not entangled.
In Fig.~1, we also observe that the maximal value of the averge linear entropy decreases
by increasing the length of the system, which agrees with the analytical results.
Moreover, 
we make numerical caculations of the ground-state linear entropy $E_s$, and
the results show a critical behavior around $\lambda =1$, which confirms the relation~(\ref{mainresult1}).

\begin{figure}
\includegraphics[width=0.40\textwidth]{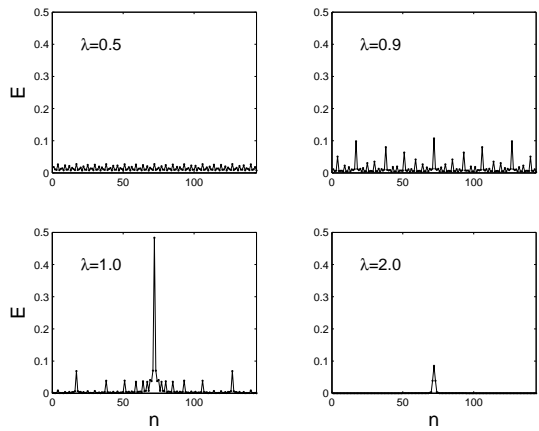}
\caption{The distribution of the linear entropy for different $\lambda$ with $N=144$
and $\sigma=89/144$.}
\end{figure}

To understand the underlying mechanism of the transition of entanglment induced by the 
change of on-site potential, we study the
distribution of the linear entropy $E_{1,N-1}^{(n)}$.
If the on-site potential is absent, the entanglment distributes evenly on the lattice
sites. 
Figure 2 shows the distribution of the linear entropy for different $\lambda$. 
When $\lambda=0.5$ smaller than $\lambda_{c}$, the entanglment
is almost even. 
When $\lambda=0.9$ near to $\lambda_c$, the uneven distribution becomes distinct but 
the state is still not localized. 
When $\lambda
=\lambda_c$, a significant change takes place, and a large peak of entanglement
appears at the center of the lattice and the entanglement at most of the other sites are suppressed.
The entanglement between the LFM at the center and the rest becomes dominant.
The ground state is localized by the effects of 
on-site potential $V_n$ in this case. 
When $\lambda=2$ larger than $\lambda_c$, the entanglment at most sites is suppressed to
zero except a few sites near the center of lattice is still of finite
value, although they are also suppressed. 
Thus, the localization is enhanced when the amplitude of 
the potential increases.
Further more, the biparticle entanglment between a block of LFMs and the
rest for such a one-particle state are
also calculated and shows similar behaviors as the linear entropy
$E_{1,N-1}$.

Having studied ground-state entanglement of the Harper model, we now
examine dynamics of entanglement.
The time evolution can be described by the following
time-dependent equation 
\begin{equation}
i\frac{d\psi _{n}}{dt}=\frac{1}{2}(\psi_{n+1}+\psi_{n-1})+\lambda \cos (2\pi n\sigma).
\end{equation}%
The wave packet is localized initially at the center of the chain and the above equation
can be solved numerically by integration. 
In our calculations, we adopt the periodic boundary condition.
The variance of the wave packet 
\begin{equation}
\sigma ^{2}(t)=\sum_{n=1}^{N}(n-\bar{n})^{2}|\psi_{n}(t)|^{2}
\end{equation}
is studied in Ref.~\cite{Gei} and different time evolution behaviors were shown.

When the state vectors at any time are obtained by integration, we can calculate  
bipartite entanglment between the block of LFMs and the rest of chain.  
In Fig.~3, we plot the average entanglment $E_{1,N-1}$ 
for several value of $\lambda$.
When $\lambda <\lambda_{c},$ $E_{1,N-1}$ exhbits a rapid initial increase, which
corresponds to the variance $\sigma^{2}(t) \sim t^{2}$~\cite{Gei}.
This unbounded diffusion is caused by the existence of extended states. In the
critical case $\lambda =\lambda _{c},$ $E_{1,N-1}$ increases
slowly, which is in agree with the clear-cut diffusion with the variance $\sigma
^{2}(t)\sim t^{1}$. For $\lambda <\lambda _{c},$ $E_{1,N-1}$ exhibits rapid
oscillations because of the localization.

\begin{figure}
\includegraphics[width=0.40\textwidth]{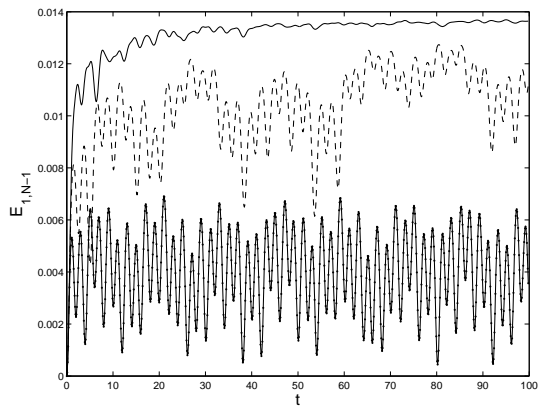}
\caption{Dynamics of the linear entropy for $\lambda=1/2$ (solid line), $\lambda=1$ (dashed line), and
$\lambda=1.5$ (connected dotted line).  The parameters $N=144$ and $\sigma=89/144$.}
\end{figure}

In this paper, we have studied bipartite entanglment of one-particle states,
and found a direct connection between the linear 
entropy, quantifying the bipartite entanglement, and the participation ratio, characterizing
state localization. The more localized the state is, the less the entanglement. 
Pairwise entanglement, bipartite entanglement, and state localization
are found closely connected together for one-particle states.

As an application of the general formalism, we have studied quantum entanglement
of the ground state in the Harper model and find that the bipartite entanglement exhibits 
a transition at critical point $\lambda=\lambda_c$. The time evolution of entanglement 
was also investigated, which displays distinct behaviors for different amplitudes of the 
on-site potential, corresponding to extended, critical, and localized states.

\acknowledgments We acknowledge valuable discussions with L. Yang and X.W. Hou.
This work was supported by the grants from the Hong Kong Research Grants Council (RGC) and the Hong Kong
Baptist University Faculty Research Grant (FRG). X. Wang has been supported by an Australian 
Research Council Large Grant and Macquarie University Research Fellowship.

\end{document}